\documentclass[11pt,a4paper]{article}

\usepackage{mathpple}
\usepackage{graphicx}
\usepackage{url}
\usepackage{paralist}

\usepackage{geometry}
\usepackage{fancyhdr}
\begin{document}

\thispagestyle{plain}

\noindent \textbf{Preprint of:}\\
T. A. Nieminen and D. K. Gramotnev\\
``Rigorous analysis of extremely asymmetrical scattering
 of electromagnetic waves in slanted periodic gratings''\\
\textit{Optics Communications} \textbf{189}, 175--186 (2003)

\hrulefill

\begin{center}

\textbf{\LARGE
Rigorous analysis of extremely asymmetrical scattering
 of electromagnetic waves in slanted periodic gratings
}

{\Large
T. A. Nieminen and D. K. Gramotnev}

Centre for Medical and Health Physics,
School of Physical Sciences,
Queensland University of Technology,
GPO Box 2434, Brisbane, QLD 4001, Australia

\vspace{1cm}

\begin{minipage}{0.8\columnwidth}
\section*{Abstract}
Extremely asymmetrical scattering (EAS) is a new type
of Bragg scattering in thick, slanted, periodic gratings. It is
realised when the scattered wave propagates parallel to
the front boundary of the grating. Its most important feature is
the strong resonant increase in the scattered wave amplitude
compared to the amplitude of the incident wave: the
smaller the grating amplitude, the larger the amplitude of
the scattered wave. In this paper, rigorous numerical analysis
of EAS is carried out by means of the enhanced T-matrix algorithm.
This includes investigation of harmonic generation
inside and outside the grating, unusually strong edge effects,
fast oscillations of the incident wave amplitude in the
grating, etc. Comparison with the previously developed
approximate theory is carried out. In particular, it is
demonstrated that the applicability conditions for the
two-wave approximation in the case of EAS are noticeably more
restrictive than those for the conventional Bragg scattering.
At the same time, it is shown that the approximate theory is
usually highly accurate in terms of description of EAS in the
most interesting cases of scattering with strong resonant
increase of the scattered wave amplitude. Physical
explanation of the predicted effects is presented.
\end{minipage}

\end{center}

\section{Introduction}

   Scattering of bulk and guided electromagnetic
waves in periodic gratings has been extensively
investigated theoretically and numerically by many
different authors using approximate and rigorous
methods of analysis [1--17]. The investigation has
been carried out for thin and thick, uniform and
non-uniform, isotropic and anisotropic, reflecting
and transmitting gratings that are represented by
strong or weak periodic modulation of structural
parameters [1--17].

   Very interesting physical effects and anomalies
of scattering have been predicted and observed at
extreme angles of scattering, i.e., when the
scattered wave propagates parallel to the grating. For
example, these are resonant coupling of bulk and
surface [16,19], or bulk and guided waves [20],
non-linear stimulated scattering involving interaction
of bulk and surface waves [21--23], Wood's
anomalies [16,17], etc. However, all these effects
and anomalies are relevant to scattering in thin
gratings, where the incident wave and/or scattered
wave interact with the grating within a short
distance of the order of, or less than the wavelength
[16--23]. For example, this happens in the case of
diffraction of a bulk electromagnetic wave on a
periodic corrugation of an interface between two
media.

    Bragg scattering in wide, oblique, periodic
gratings with the scattered wave propagating parallel
to grating boundaries has been investigated less
extensively compared to thin gratings. In the
beginning of 1970s a radically new type of Bragg
scattering of X-rays in crystals and crystal plates
(i.e. thick gratings) was analysed theoretically [24--27].
It was called extremely asymmetrical scattering (EAS).
However, the main efforts in theoretical
and experimental investigation of EAS of X-ray
and neutrons were focused on the cases where an
incident wave propagates almost parallel to the
front boundary of a crystal [28]. This allowed very
precise structural analysis of interfaces and
ultra-thin crystal films, and resulted in the development
of efficient collimators of X-rays and neutron
beams [28]. At the same time, the geometry of EAS
with grazing incident wave is not that interesting
for the development of applications in optical
communication and instrumentation. Much more
important is the geometry where the incident wave
propagates at a significant angle with respect to
boundaries of a strip-like oblique periodic grating,
and the scattered wave is parallel or almost parallel
to these boundaries. This is because of unique
features displayed by EAS in this geometry [29--37].
However, theoretical analysis of EAS of bulk
and guided optical and surface acoustic waves has
been carried out only within the last few years [29--37].

   The diffractional divergence of the scattered
wave inside and outside the grating has been
shown to be the main physical reason for EAS
[29--31,33,34]. A new powerful approach for the
analytical analysis of EAS, based on understanding
the role of the diffractional divergence, has been
developed and justified [29--31,33,34]. The most
important feature of this approach is that it is
immediately applicable for the description of EAS
of all types of waves (including bulk, guided and
surface optical and acoustic waves) in all types of
periodic gratings with small amplitude [29--37].

   It has been shown that EAS is characterised by
a resonantly large scattered wave amplitude (the
most interesting case of scattering) only if the
grating amplitude is very small [29--34]. In this
case, the applicability conditions for the new
approach are usually well satisfied [37], and the
analytical theory is expected to be accurate in
predicting amplitudes of the incident and
especially scattered waves.

   Since the main direction in the development of
the modern theory of gratings is related to
improvement of stability and convergence speed of
numerical algorithms (see for example Ref. [9]),
the development of the new numerically efficient
(analytical) method for the accurate description of
strong EAS (of all types of waves, including guided
and surface modes) is an important step in the
grating theory. Moreover, this method has provided
a unique insight into the physical reasons for
EAS [29--37]. This insight will allow thoughtful
selection of optimal structural parameters for
future EAS-based devices and techniques.

   Nevertheless, despite the fact that the new
analytical approach is expected to describe accurately
EAS with strong resonant increase of the scattered
wave amplitude, it is still unknown what happens
to EAS beyond the frames of the applicability of
the analytical approach. That is, what changes in
the pattern of scattering should be introduced if the
grating amplitude is strongly increased and/or
grating width is significantly reduced? What are
the exact errors of using the analytical approach
for various grating amplitudes and widths? How
accurate are the applicability conditions for the
approximate theory? What are the most crucial
parameters affecting the applicability of the
analytical approach? Which of the two waves---%
incident or scattered---is better described by the
approximate theory? All these questions remained
largely unanswered in the previous publications.

   Therefore, the main aim of this paper is to
implement detailed rigorous analysis of EAS of bulk
TE electromagnetic waves in narrow and wide
periodic gratings of arbitrary amplitude, represented
by sinusoidal variations of the dielectric
permittivity. In addition, the analytical
applicability conditions for the approximate theory of
EAS will be verified and compared with the results
of the rigorous theory. Physical mechanisms that
are responsible for breaching these conditions will
be discussed. Typical errors related to the
approximations of the analytical approach will be
determined.

\section{Numerical analysis}

    Consider an isotropic medium with a slab-like
holographic periodic grating that is characterised
by sinusoidal variations of the dielectric permittivity:
\begin{eqnarray}
\epsilon_s & = & \epsilon +
(\epsilon_1 \exp(\mathrm{i}q_xx + \mathrm{i}q_yy) + \textrm{c.c.}),
\,\,\,\, \textrm{if } 0<x<L, \nonumber \\
\epsilon_s & = & \epsilon, \,\,\,\, \textrm{if }
x<0 \textrm{ or } x>L.
\end{eqnarray}
where the coordinate system is shown in Fig. 1, $\epsilon$ is
the mean dielectric permittivity that is the same
inside and outside the grating, $\epsilon_1$ is the complex
amplitude, $\mathbf{q} = (q_x,q_y)$ is the reciprocal lattice
vector, $q = 2\pi/\Lambda$, $\Lambda$ is the period, and $L$ is the
width of the grating that is assumed to be infinite
along the $y$- and $z$-axes. We also assume that the
dissipation is absent, i.e., $\epsilon$ is real and positive. A
TE electromagnetic wave with the amplitude $E_{00}$
and wave vector $\mathbf{k}_0$ is incident onto the grating at
an angle $\theta_0$ in the $xy$  plane---Fig. 1 (non-conical
scattering).

\begin{figure}[!h]
\centerline{\includegraphics[width=0.5\columnwidth]{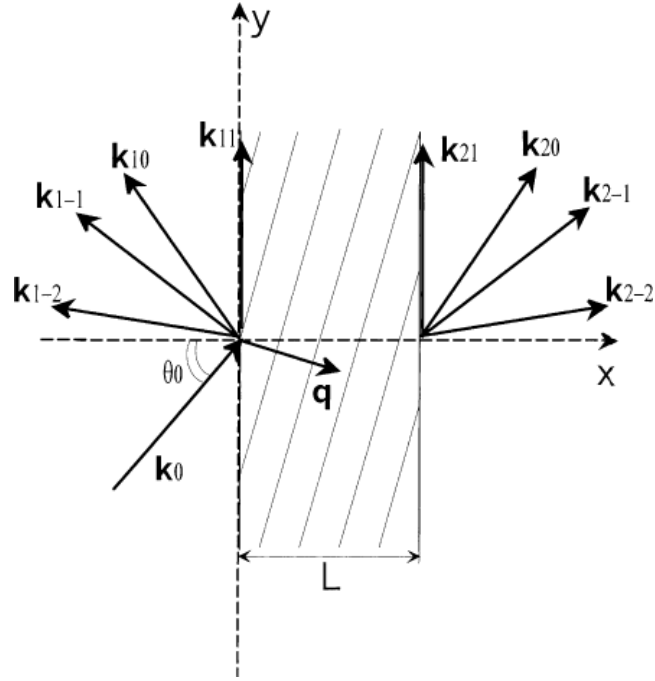}}
\caption{The geometry of EAS in a slanted periodic grating of
width $L$. The Bragg condition is satisfied precisely for the $+1$
harmonic that propagates parallel to the grating boundaries
$x=0$ and $x=L$. The angle of incidence $\theta_0$ is not close to $\pi/2$.}
\end{figure}

   In this case the solutions inside and outside the
grating can be written as [7,8]
\begin{equation}
E(x,y) = \sum_{h=-\infty}^{+\infty} E_h(x)
\exp(\mathrm{i}xk_{hx} + \mathrm{i}yk_{hy}),
\end{equation}
\begin{equation}
E|_{x<0} = E_{00}\exp(\mathrm{i}\mathbf{k}_0\cdot\mathbf{r})
+ \sum_{h=-\infty}^{+\infty} A_h
\exp((\mathrm{i}\mathbf{k}_{1h}\cdot\mathbf{r}),
\end{equation}
\begin{equation}
E|_{x>L} = \sum_{h=-\infty}^{+\infty} B_h
\exp(\mathrm{i}\mathbf{k}_{2h}\cdot\mathbf{r} - \mathrm{i}Lk_{2hx}),
\end{equation}
where $k_{hx}$ and $k_{hy}$ are the $x$- and $y$-components of
the wave vectors
\begin{equation}
\mathbf{k}_h = \mathbf{k}_0 - h\mathbf{q}
\,\,\,\, (h = 0; \pm 1, \pm 2, ... ),
\end{equation}
the components of the wave vectors k1 h and k2 h are
determined by the equations:
\begin{equation}
k_{1hy} = k_{2hy} = k_{hy}, \,\,\,\,
k_{1hx} = -k_{2hx} = -(k_0^2 - k_{hy}^2)^{1/2}.  
\end{equation}
If $k_{hy} \le k_0$, then $k_{1hx} \le 0$ and
$k_{2hx} \ge 0$ (propagating
waves), while if $k_{hy} > k_0$, then $\mathrm{Im}(k_{1hx}) < 0$ and
$\mathrm{Im}(k_{2hx}) > 0$ (evanescent waves).

   The Bragg condition is assumed to be satisfied
precisely for the $+1$ harmonic in Eq. (2), i.e. for the
first diffraction order with $h=1$. In addition, the
$+1$ harmonic is assumed to propagate parallel to
the $x$-axis (the geometry of EAS).

\begin{figure}[!t]
\centerline{\includegraphics[width=0.7\columnwidth]{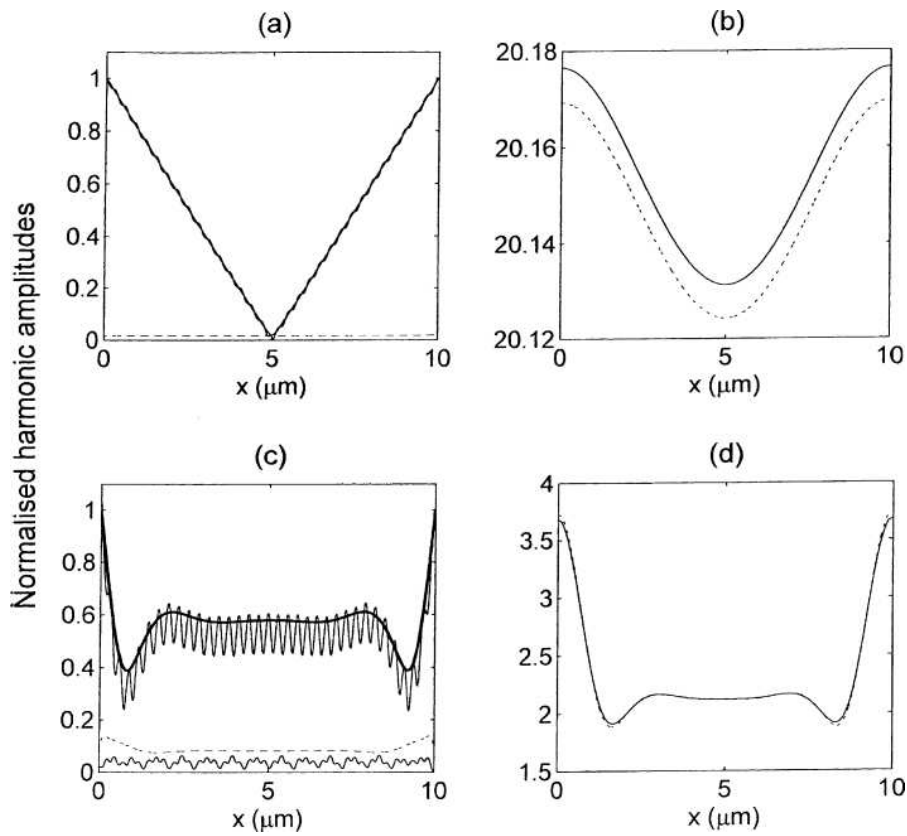}}
\caption{The $x$-dependencies of the normalised amplitudes
$|E_h/E_{00}|$ of harmonics in Eq. (2) inside the grating
with the parameters: $\epsilon=5$, $\theta_0=45^\circ$,
$L=10\mu$, $\lambda=1\mu$ (in vacuum), and with the grating
amplitudes: (a,b) $\epsilon_1 = 5\times 10^{-3}$,
(c,d) $\epsilon_1 =0.2$. The grating orientation and
period $\Lambda \approx 0.584\mu$m are determined by
the Bragg condition. (a) The rigorous (solid curve
with small oscillations) and
approximate (solid curve without oscillations) $x$-dependencies
of the amplitude of the zeroth harmonic (incident wave) inside the
grating. Dotted curve---the rigorous $x$-dependence of the $+2$
harmonic amplitude. (b,d) The rigorous (dotted curves) and approximate
(solid curves) dependencies of amplitudes of the $+1$
harmonics (the scattered waves) inside the grating.
(c) The thick solid curve is the
approximate $x$-dependence of the amplitude of the incident
wave inside the grating. The rigorous $x$-dependencies
of amplitudes of the
0th, $+2$, and $-1$ harmonics are presented by
the higher thin solid curve, dotted curve, and
lower thin solid curve, respectively.}
\end{figure}

\begin{figure}[!t]
\centerline{\includegraphics[width=0.7\columnwidth]{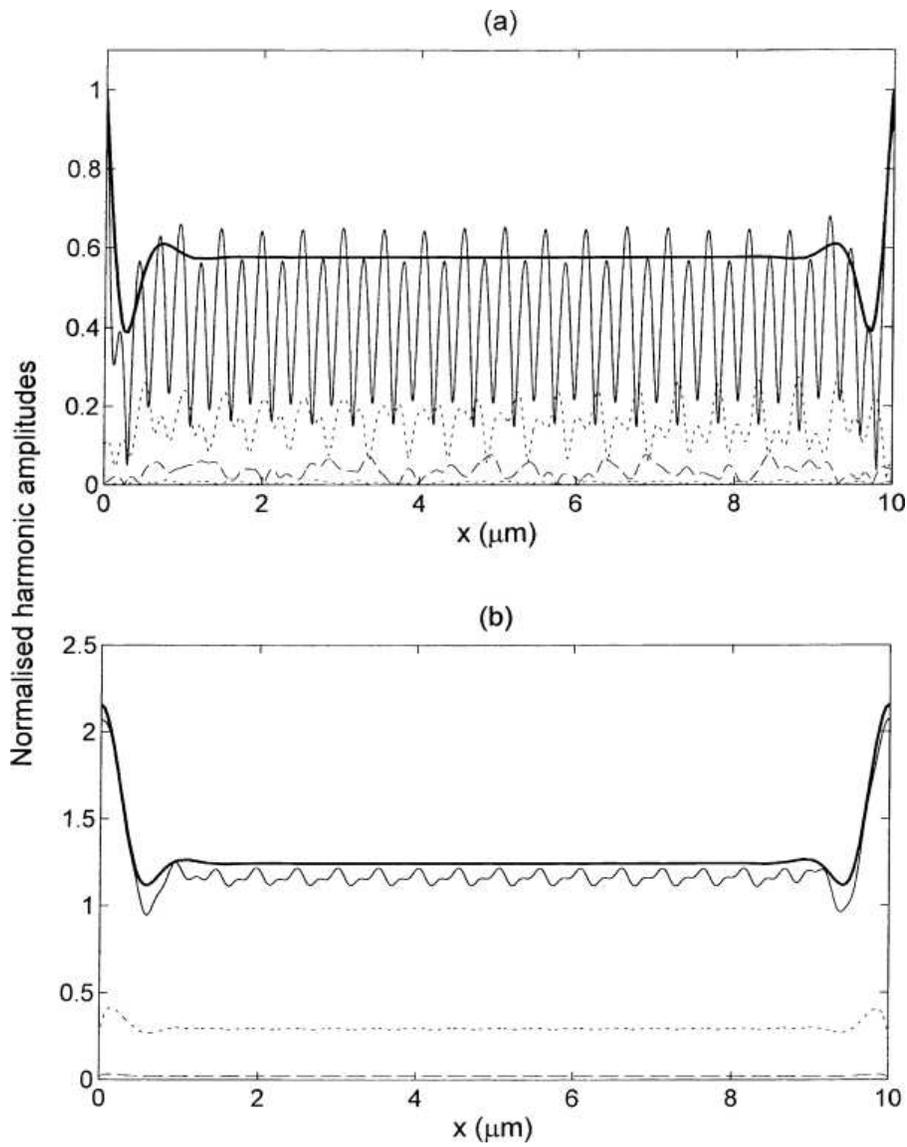}}
\caption{The rigorously calculated $x$-dependencies of the
normalised amplitudes $|E_h/E_{00}|$ of harmonics in Eq. (2)
in the same structure as
for Fig. 2, but with $\epsilon_1 =1$. (a) The rigorous
dependencies of amplitudes of the 0th harmonic (thin solid curve),
$-1$ harmonic (higher
dotted curve), $-2$ harmonic (dashed curve), and $-3$
harmonic (lower dotted curve). (b) The rigorous dependencies
of amplitudes of the
$+1$ harmonic (thin solid curve), $+2$ harmonic (dotted curve),
and $+3$ harmonic (dashed curve). The thick solid curves represent the
approximate $x$-dependencies of the incident (a) and scattered (b)
wave amplitudes, obtained from the approximate theory [29--32].}
\end{figure}

   The rigorous analysis of first-order scattering in
this geometry has been carried out by means of
the enhanced transmittance matrix algorithm [7].
Fig. 2 presents the approximate and rigorous
$x$-dependencies of normalised amplitudes of
harmonics from Eq. (2) inside the grating for EAS of
bulk TE electromagnetic waves. The structural
parameters are as follows:
$\epsilon=5$, $\theta_0=\pi/4$, $L=10\mu$m,
vacuum wavelength $\lambda=1\mu$m, and the grating
amplitudes are $\epsilon_1 = 5\times 10^{-3}$ (Fig. 2a and b)
and $\epsilon_1 =0.2$ (Fig. 2c and d). The grating period is
determined by the Bragg condition: $\Lambda \approx 0.584\mu$m.
Similar dependencies in Fig. 3a and b are presented
for the same structure but with the grating
amplitude $\epsilon_1 =0.2$. Only harmonics whose
amplitudes are larger than $\approx 0.01E_{00}$ are presented in
Figs. 2 and 3.

   It can be seen that in gratings with small
amplitudes, where the scattered wave amplitude is
especially large (Fig. 2a and b), the eect of all
harmonics in the expansion (2), other than the
zeroth harmonic (the incident wave) and the $+1$
harmonic (the scattered wave), is completely
negligible. The rigorous $x$-dependencies of the
incident and scattered wave amplitudes inside the
grating hardly dier from the corresponding
approximate dependencies that have been
determined by means of the approximate theory [31]
(Fig. 2a and b). Thus, for sufficiently small grating
amplitudes the approximate theory [29--34]
provides a very accurate description of EAS and there
is no need for the rigorous analysis. The only other
harmonic (in addition to the zeroth and $+1$
harmonics) that may have reasonably noticeable
amplitude is the $+2$ harmonic. This is due to the
direct coupling of the resonantly strong scattered
wave (the $+1$ harmonic) and the $+2$ harmonics [4].

   However, if the grating amplitude is significantly
increased from $5\times 10^{-3}$ to $0.2$ (Fig. 2c and
d) and then to 1 (Fig. 3a and b), then the
approximate theory is not always sufficient for the
description of EAS. This is especially the case for
the amplitude of the incident wave, i.e. the zeroth
harmonic in Eq. (2). For example, if $\epsilon_1 =0.2$ (Fig.
2c), then the rigorously calculated $x$-dependence of
the zeroth harmonic amplitude is noticeably lower
(on average) than the corresponding approximate
dependence (Fig. 2c), and is characterised by a
number of oscillations (see also Fig. 3a).

   These oscillations are related to boundary
scattering of the scattered wave at the grating
interface $x=L$. The wave resulting from boundary
scattering of the scattered wave at the rear
boundary $x=L$ propagates in the negative
$x$-direction as if it is a mirror reflected incident wave
from this boundary. Thus, the $y$-component of its
wave vector is equal to the $y$-component of the wave
vector of the incident wave. However, the wave
due to boundary scattering is not presented
explicitly in Eq. (2). Therefore the zeroth harmonic
in Eq. (2) includes both the incident wave and the
wave caused by boundary scattering at $x=L$. The
interference of these waves results in a standing
wave pattern represented by the fast oscillations of
the zeroth harmonic amplitude---Figs. 2c and 3a.
It can be seen that the period of this standing wave
pattern must be equal to $\lambda/(2\epsilon^{1/2}\cos\theta_0)$,
which is
in the excellent agreement with the rigorous
dependencies in Figs. 2c and 3a.

   As mentioned above, the average (over the period
$\lambda/(2\epsilon^{1/2}\cos\theta_0)$)
rigorously calculated amplitude of the
zeroth harmonic tends to be smaller
than the amplitude predicted by the approximate
theory---see Figs. 2c and 3a. The larger the grating
amplitude, the smaller the average amplitude of
the zeroth harmonic inside the grating (Figs. 2c
and 3a). This is related to significant boundary
scattering at the front grating interface, which
results in energy losses in the zeroth harmonic
with subsequent reduction of its amplitude in the
grating.

   The contribution of higher harmonics to
scattering rapidly increases when the grating
amplitude $\epsilon_1$ exceeds $\approx 10$\%
of the mean permittivity $\epsilon$
(compare Figs. 2c and 3a,b). It is however
important that amplitude of the scattered wave is
accurately described by the approximate theory
(within an error less than $\approx 1$\%) up to grating
amplitudes such that $\epsilon_1/\epsilon \approx 0.1$.
Only when $\epsilon_1/\epsilon = 0.2$ can
noticeable deviations between the rigorous
and approximate amplitudes of the $+1$ harmonic
be observed (Fig. 3a and b).

   Increasing grating amplitude should result not
only in noticeable amplitudes of several harmonics
in Eq. (2) (see Figs. 2 and 3), but also in significant
energy flows from the grating due to propagating
waves in Eqs. (3) and (4) (in the considered
examples these will correspond to $h= -1,..,-5$).
In addition, the zeroth harmonic in the sum of Eq.
(3), caused by boundary scattering at the grating
interfaces, also results in an energy flow away from
the grating in the region $x < 0$. The associated
energy losses can be evaluated in terms of diffraction
efficiencies that are determined as ratios of the
$x$-component of the Poynting vector in a wave
travelling away from the grating to the $x$-component
of the Poynting vector in the incident wave at
$x < 0$. If the diffraction efficiencies are small
compared to one, the energy losses are negligible, and
the approximate theory of EAS is valid.

\begin{figure}[htb]
\centerline{\includegraphics[width=0.7\columnwidth]{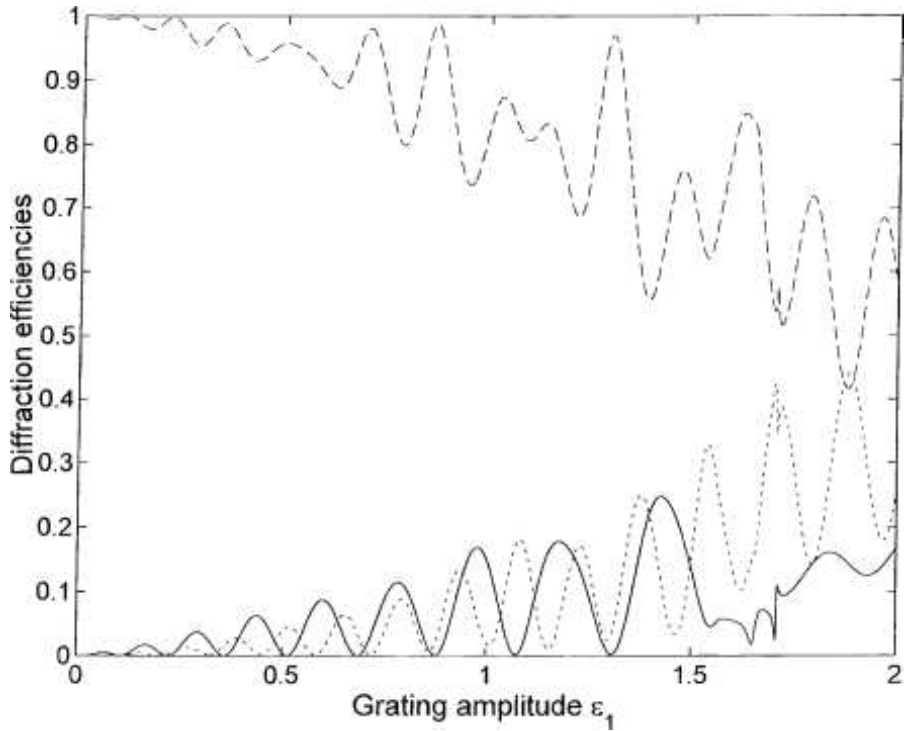}}
\caption{The dependencies of the diffraction efficiencies on
grating amplitude $\epsilon_1$ for the transmitted
wave with the amplitude $B_0$
(dashed curve), zeroth reflected wave with the amplitude $A_0$
(solid curve), and the combined efficiency for all other
propagating waves outside the grating (dotted curve). The other
structural parameters are as for the previous figures:
$\epsilon=5$, $\theta_0=\pi/4$, $L=10\mu$m,
$\lambda=1\mu$m, and $\Lambda \approx 0.584\mu$m.}
\end{figure}

   The dependencies of the diffraction efficiencies
in the structure with $\epsilon=5$, $\theta_0=\pi/4$,
$L=10\mu$m, and $\lambda=1\mu$m on grating amplitude
$\epsilon_1$ are presented in Fig. 4. The dashed curve represents the
diffraction efficiency of the transmitted wave, i.e.
the zeroth harmonic in Eq. (4). The solid line gives
the diffraction efficiency of the wave resulting from
boundary scattering (i.e. the zeroth reflected wave
for $x < 0$), and the dotted curve gives the overall
diffraction efficiency from all other waves
propagating outside the grating. Note that the
diffraction efficiency for the scattered wave, i.e.
the $+1$
harmonic, is zero because it propagates parallel to
the grating boundaries in both the regions $x < 0$
and $x > L$.

   It can be seen that if the grating amplitude
$\epsilon_1 < 0.12$ (i.e. $\epsilon_1 < 0.025\epsilon$),
then all the diffraction efficiencies other
than that of the transmitted
wave are less than $\approx 0.01$ and can be neglected.
The diffraction efficiency for the transmitted wave
in this case is $\approx 1$ (Fig. 4).

   It is interesting that all the diffraction
efficiencies experience significant oscillations with
increasing grating amplitude e1 . This is due to
the interference of waves at the grating boundaries.
For example, the waves caused by boundary
scattering at the front and rear grating
boundaries interfere constructively or destructively
in the region $x < 0$, resulting in the oscillations of
the solid curve in Fig. 4.

    All these results demonstrate that in the most
interesting case of EAS with strong resonant
increase of the scattered wave amplitude the
approximate theory [29--34] gives very accurate
results, especially for the scattered wave amplitude.
Only when the grating amplitude increases so
that the resonant scattered wave amplitude
becomes of the order of $E_{00}$, does the approximate
theory fail to accurately predict, first, the incident
wave amplitude inside the grating, and only then
(if the grating amplitude is increased up to
$\approx 0.2\epsilon$)
the scattered wave amplitude. Note, however, that
this is correct only for gratings of widths that are
not much less than the critical grating width $L_c$
[34--37]
\begin{equation}
L_c \approx (2c/\omega)[2\epsilon^{-1}|A_1/(E_{00}\epsilon_1)|]^{1/2},
\end{equation}
where $A_1$ is the amplitude of the $+1$ harmonic at
the front boundary in a wide grating (with $L > _Lc$---%
see Refs. [34--36]).

   If the grating width $L$ is decreased below $L_c$,
then the approximate theory predicts that the
scattered wave amplitude inside and outside the
grating must increase proportionally to $L^{-1}$
[29--32,37]. In this case, even if the grating amplitude is
small, the direct coupling of the scattered wave
(the $+1$ harmonic) to the $+2$ harmonics must
result in increasing amplitude of the $+2$ harmonic
inside the grating. In addition, the wave resulting
from boundary scattering (i.e. the zeroth harmonic
in the sum in Eq. (3)) must also increase proportionally
to the amplitude of the scattered wave, i.e.
proportionally to $L^{-1}$. These effects suggest that
decreasing grating width will eventually lead to the
breach of the applicability of the approximate
theory [29--32,37].

\begin{figure}[!t]
\centerline{\includegraphics[width=0.7\columnwidth]{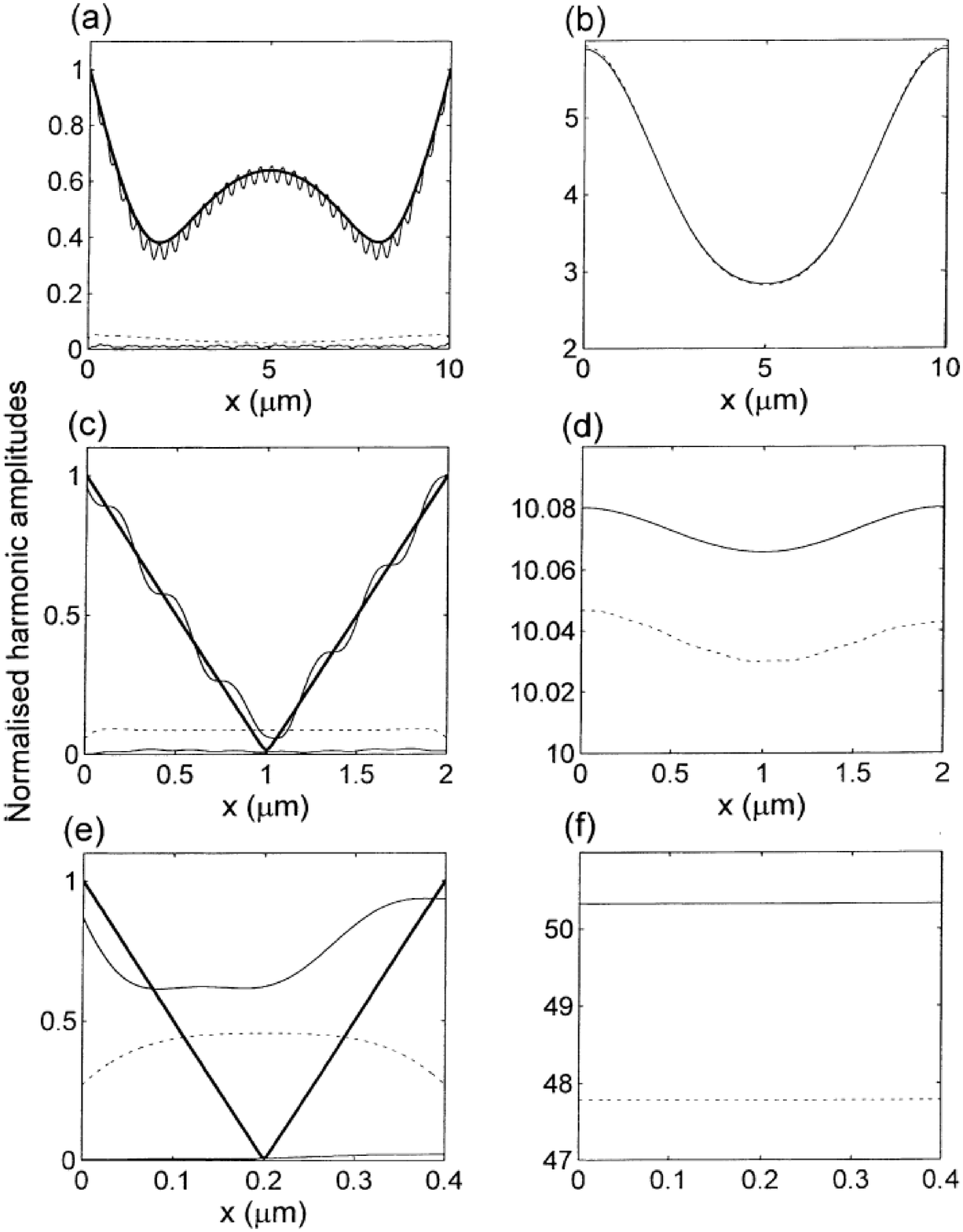}}
\caption{The $x$-dependencies of the normalised amplitudes
$|E_h/E_{00}|$ of harmonics inside the grating with the parameters:
$\epsilon=5$, $\epsilon_1 = 5\times 10^{-2}$, $\theta_0=\pi/4$,
$\lambda=1\mu$, $\Lambda \approx 0.584\mu$m, and the grating widths:
(a,b) $L=10\mu$m, (c,d) $L=2\mu$m, (e,f) $L=0.4\mu$m. (a), (c), and (e)
present the rigorous $x$-dependencies of amplitudes of the
0th harmonics (higher thin solid curves),
$+2$ harmonics (dotted curves), and
$-1$ harmonics (lower thin solid curves).
The thick solid curves are the approximate $x$-dependencies
of the incident wave amplitudes in
the gratings. (b), (d), and (f) present the
rigorous (dotted curves) and approximate (solid curves)
$x$-dependencies of amplitudes of the
$+1$ harmonics (scattered waves).}
\end{figure}

   Fig. 5 presents the results of the rigorous and
approximate analyses of EAS in gratings of different
widths: (a) and (b) $L=10\mu$m; (c) and (d)
$L=2\mu$m; (e) and (f) $L=0.4\mu$m. The other parameters are:
$\epsilon=5$, $\epsilon_1 = 5\times 10^{-2}$, $\theta_0=\pi/4$,
$\lambda=1\mu$.

   We can see that for $L=10\mu$ the pattern of
scattering is very much the same as in Fig. 2a--d.
The incident wave amplitude inside the grating
experiences oscillations similar to those in Fig. 2c,
but with smaller amplitude (due to smaller $\epsilon_1$). The
amplitudes of the $+2$ and $-1$ harmonics are small
(Fig. 5a), and the approximate and rigorous curves
for the scattered wave amplitude are practically
indistinguishable---Fig. 5b.

   If the grating width is reduced to $2\mu$m (Fig. 5c
and d), then the number of oscillations of the
rigorous dependence of the incident wave amplitude
inside the grating is significantly reduced.
This is because fewer nodes of the standing wave
pattern can fit across a narrower grating. The
amplitude of the $+2$ harmonic is increased $\approx 2$
times compared to the amplitude of the same
harmonic in Fig. 5a. This is because the amplitude
of the scattered wave in Fig. 5d is $\approx 2$ times larger
than in Fig. 5b. Note that for $L=2\mu$m, the
rigorous dependence of the $+1$ harmonic amplitude is
only $\approx 0.03$\% different from the approximate curve
(Fig. 5d). That is, the oscillations of the incident
wave amplitude and noticeable amplitude of the
2 harmonic (Fig. 5c) hardly affect the scattered
wave amplitude.

   If the grating width is decreased further down
to $0.4\mu$m, then the rigorously calculated dependence
of the incident wave amplitude in the grating
becomes drastically different from the approximate
dependence (Fig. 5e). The amplitude of the $+2$
harmonic becomes large and comparable with the
amplitude of the incident wave (Fig. 5e). Note
however, that since the $-1$ harmonic is not
coupled directly to the resonantly large amplitude of
the $+1$ harmonic, the amplitude of the $-1$
harmonic is hardly affected by reducing grating width
(compare the lower thin solid curves in Fig. 5a, c,
and e).

   For the grating width of $0.4\mu$m the $+1$
harmonic amplitude is so large (Fig. 5f), that
boundary scattering results in a significant energy flow
from the grating. This is the main reason why the
rigorously calculated amplitudes of the $+1$
harmonic (scattered wave) appear to be noticeably
smaller than those obtained in the approximate
theory [31] (Fig. 5f). The difference between these
curves is about 5\%. Thus, edge effects in the form
of boundary scattering are the main reason for
the failure of the approximate theory [29--32] to
accurately predict scattered wave amplitudes in the
case of EAS in narrow gratings.

   Note again that all harmonics in Eq. (2), other
than the zeroth, $+1$, $+2$, and $-1$ harmonics, have
very small amplitudes in the considered structures
for all grating widths (Fig. 5a--f), and thus can be
neglected.

\begin{figure}[htb]
\centerline{\includegraphics[width=0.7\columnwidth]{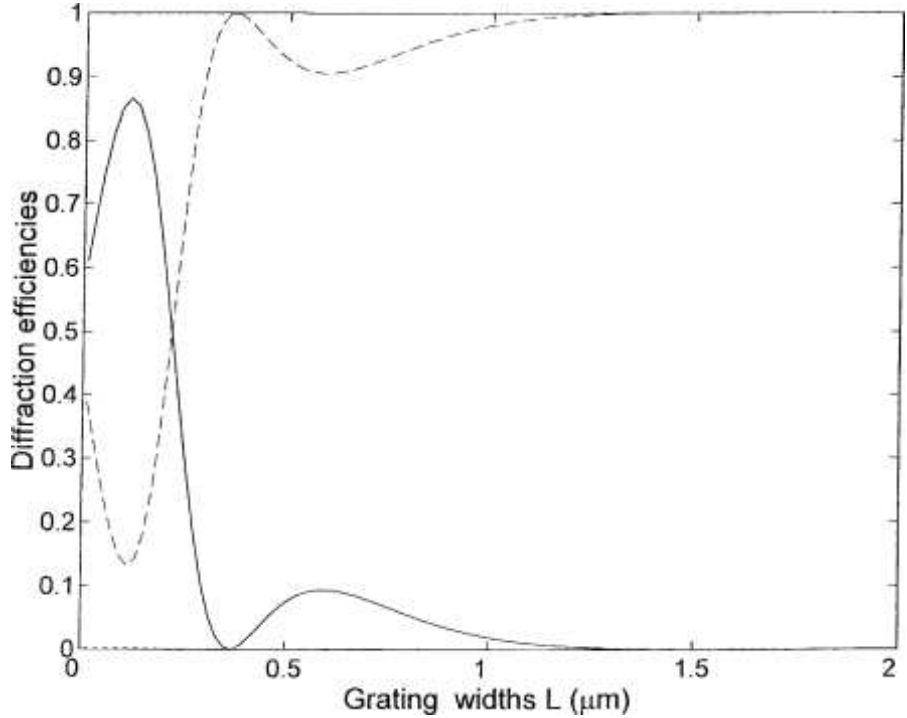}}
\caption{The dependencies of the diffraction efficiencies on
grating width $L$ for the transmitted wave with the amplitude $B_0$
(dashed curve), zeroth reflected wave with the amplitude $A_0$
(solid curve), and the combined efficiency for all other
propagating waves outside the grating (dotted curve)---can hardly be
seen near the origin of the graph. The structural parameters are
the same as for Fig. 5:
$\epsilon=5$, $\epsilon_1 = 5\times 10^{-2}$, $\theta_0=\pi/4$,
$\lambda=1\mu$, and $\Lambda \approx 0.584\mu$m.}
\end{figure}

    The diffraction efficiencies versus grating width
for the structure considered for Fig. 5 are
presented in Fig. 6. The dashed curve gives the
diffraction efficiency of the transmitted wave (the
zeroth harmonic at $x > L$), and the solid curve
gives the diffraction efficiency for the wave due to
boundary scattering. The overall efficiency for
other propagating waves is significantly less than
1\%, and the relevant dotted curve can hardly be
seen close to the origin of the graph (Fig. 6). This
figure demonstrates that, as mentioned above,
decreasing grating width below $1\mu$m results in
significant boundary scattering. This results in the
failure of the approximate theory. However, for
grating widths $L > 1\mu$m all diffraction efficiencies,
except for that of the transmitted wave, are
negligible (Fig. 6), and the approximate theory gives
very accurate results, at least in terms of predicting
scattered wave amplitudes inside and outside the
grating.

   Note that the analysis of the case with $L > L_c$
has already been carried out in Figs. 2c,d, 3a,b,
and 4. Therefore, increasing grating width beyond
$10\mu$m does not reveal any new features of EAS. In
addition, the typical number of oscillations of the
rigorously calculated dependencies of the wave
amplitudes (see Figs. 2, 3, and 5) increases
proportionally to $L$, which would make relevant
figures unreadable. Therefore, the numerical results
have been presented only for gratings of $L \le 10\mu$m,
though the enhanced T-matrix approach is
stable for gratings of arbitrary width [7,8].

\section{Applicability conditions for the approximate theory}

   As was demonstrated in the previous section,
the approximate theory of EAS, based on the
two-wave approximation and the scalar theory of
diffraction of the scattered wave [29--32], normally
gives accurate results in the most interesting cases
of scattering with strong resonant increase of the
scattered wave amplitude. It has also been shown
[37], that the applicability conditions for the
approximate theory of EAS can be evaluated in two
different ways. First, we take the applicability
condition for the two-wave approximation in the
case of the conventional Bragg scattering of TE
electromagnetic waves [4]
\begin{equation}
\rho = \lambda^2\Lambda^{-2}|\epsilon_1|^{-1}
=q^2\epsilon/(k_0^2|\epsilon_1|) > 10
\end{equation}
and divide $\rho$ by the normalised maximal value of
the scattered wave amplitude:
\begin{equation}
\rho_\mathrm{EAS} = \rho|E_{00}/\mathrm{max}(|E_1(x)|)| > 10
\end{equation}
This inequality can be regarded as the applicability
condition for the approximate theory of EAS
[37]. The factor $|E_{00}/\mathrm{max}(|E_1(x)|)|$ appears in
inequality (9) because in the case of EAS the
scattered wave amplitude is much larger than the
amplitude of the incident wave: $|E_1| \ll |E_{00}|$. As
indicated in Section 2, this may result in unusually
strong boundary scattering, oscillations of the
incident wave amplitude inside the grating, and
large amplitude of the 2 harmonic (all these
effects are proportional to $|E_1|$).

   If inequality (12) is satisfied, then the errors in
the energy flux in the scattered wave, which result
from the use of the two-wave approximation, are
expected to be of the order of $\approx 1/\rho_\mathrm{EAS}^2$,
i.e., less than 1\% (see also Ref. [4]).

   Another way of evaluating applicability conditions
for the approximate theory [29--32,37] is to
directly evaluate boundary scattering [37]. Figs. 4
and 6 demonstrate that boundary scattering of a
resonantly large scattered wave amplitude is the
main source of error in the case of small grating
amplitudes. For large grating amplitudes, it gives
approximately the same average diffraction
efficiency as all other harmonics with $h\ne 0,1$ (Fig. 4).
Therefore, conditions for neglecting boundary
scattering can be regarded as the applicability
conditions for the approximate theory of EAS
[29--32,37]. The evaluation of the efficiency of
boundary scattering has demonstrated [37] that the
energy flow in the zeroth reflected wave at $x < 0$
can be neglected if
\begin{eqnarray}
4(\Delta x/L)^2 & \ll & 1 \,\,\,\, \textrm{if } L\le L_c,
\nonumber \\
4(\Delta x/L_c)^2 & \ll & 1 \,\,\,\, \textrm{if } L>L_c,
\end{eqnarray}
where $L_c$ is the critical grating width determined
by Eq. (7) [34--37], and $\Delta x$ is the typical thickness
of the region around the front grating boundary
[37], where from the energy of the scattered wave is
transferred into the energy of the boundary scattered wave.

   If conditions (10) are satisfied, then
the diffraction efficiency for the boundary
scattered wave in the region $x < 0$ should be of the
order of the left-hand sides of inequalities (10)
[37]. Knowing these efficiencies, we can
easily evaluate typical variations of the scattered
wave amplitude inside the grating.

   If $L < L_c$, then in accordance with condition
(10), the diffraction efficiency for the transmitted
wave with the amplitude $B_0$ (see Eq. (4)) can be
evaluated as
\begin{displaymath}
|B_0|^2/|E_{00}|^2 \approx 1-r,
\end{displaymath}
where $r = 4(\Delta x/L)^2$. From this equation we have:
\begin{equation}
|B_0| \approx |E_{00}|(1-r/2).
\end{equation}
If $L < L_c$ , then the approximate theory [29--32]
gives that the amplitude of the incident wave inside
the grating reduces linearly from the magnitude
$|E_{00}|$ to approximately zero in the middle of the
grating, and then increases linearly back to
$|E_{00}|$---see Figs. 2a, 5c and e. This is due to re-scattering
of the scattered wave inside the grating [36,37].
The amplitude of the re-scattered wave increases
linearly with increasing $x$ in the grating and has a
phase shift $\approx\pi$ with respect to the incident wave.
In the middle of the grating, the magnitude of the
re-scattered wave amplitude becomes equal to
$|E_{00}|$, and at the rear grating boundary $x=L$ it
reaches approximately $2|E_{00}|$. If there is noticeable
boundary scattering, then the amplitude of the
re-scattered wave at the rear boundary is less than
$2|E_{00}|$ by the value $\Delta E_0\approx r|E_{00}|/2$
(see Eq. (11)).
On the other hand, the rate of increasing re-scattered
wave amplitude along the $x$-axis is directly
proportional to the scattered wave amplitude.
Thus we can write:
\begin{equation}
2|E_{00}| - r|E_{00}|/2 \approx G|E_1|L,
\end{equation}
where $G$ is the coefficient of proportionality
between $|E_1|$ and the rate of increasing re-scattered
wave amplitude in the grating. The amplitude of
the scattered wave inside the grating is reduced due
to energy losses caused by boundary scattering.
Thus, $|E_1|=|E_{10}|-\Delta E_1$, where $E_{10}$
is the scattered
wave amplitude determined by means of the approximate
theory [29--32], and $\Delta E_1$ is the error in
the magnitude of this amplitude. Substituting this
equation for $|E_1|$ in Eq. (12), and taking into
account that $A|E_{10}|L \approx 2|E_{00}|$, we obtain:
\begin{equation}
\Delta E_1 \approx r|E_{00}|/(2AL)\approx
2\Delta x^2|E_{00}|/(AL^3).
\end{equation}
Using Eq. (12) again, we get:
\begin{eqnarray}
\Delta E_1/|E_1| & \approx & (\Delta x/L)^2,
\,\,\,\, \textrm{if } L \le L_c,
\nonumber \\
\Delta E_1/|E_1| & \approx & (\Delta x/L_c)^2,
\,\,\,\, \textrm{if } L > L_c,
\end{eqnarray}

   It is obvious that $\Delta x$ should be of the order of,
or less than one wavelength in the medium. More
accurate evaluation of the thickness $\Delta x$ can be
obtained from the comparison of conditions (14)
with the rigorous numerical results.

\begin{figure}[htb]
\centerline{\includegraphics[width=0.7\columnwidth]{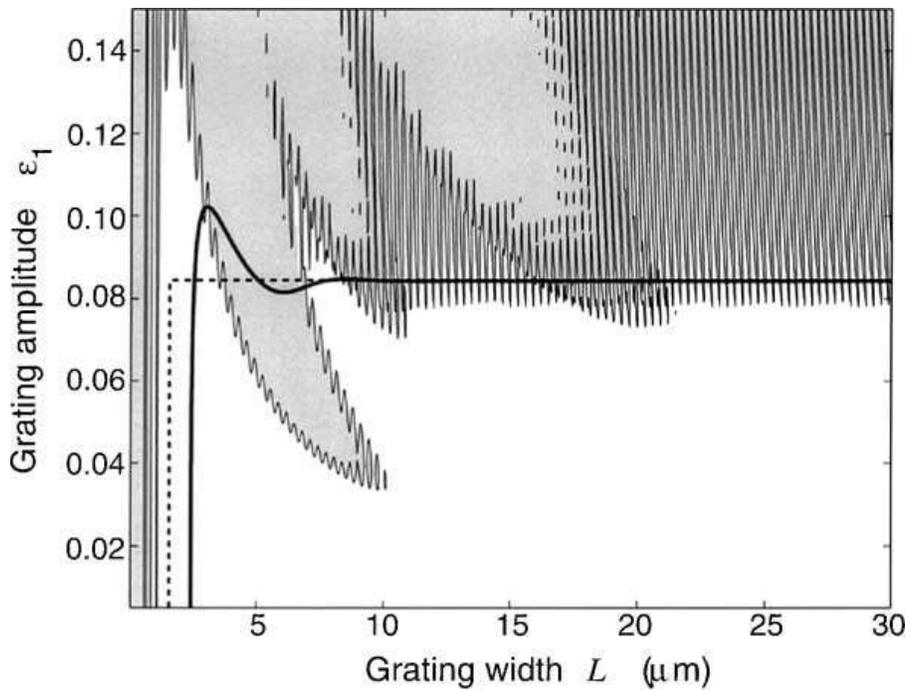}}
\caption{The rigorously calculated contours of 1\% difference
between the approximate and rigorous scattered wave amplitudes.
In the shaded regions, the maximal error in the approximate
scattered wave amplitude are larger than 1\%,
whereas in the unshaded regions this error are less than 1\%. The
dashed rectangle is the approximate contour of 1\% error in the
scattered wave amplitude, determined by Eqs. (14)
with $\Delta x \approx \lambda_m/\mathrm{e}$.
Inside this rectangle, the error in the approximate
scattered wave amplitude is expected to be smaller than
1\%. The thick solid curve represents the approximate contour
of 1\% error in the scattered wave amplitude, determined from
Eq. (15).}
\end{figure}

   Fig. 7 presents the contour plot for the 1\% error
in the approximate scattered wave amplitude [31]
as a function of grating width L and grating
amplitude $\epsilon_1$ for EAS of bulk TE electromagnetic
waves. The structural parameters are as previously:
$\epsilon=5$, $\theta_0=\pi/4$,
$\lambda=1\mu$, and $\Lambda \approx 0.584\mu$m.
In the shaded regions the maximal difference
(error) between the rigorous and approximate
$x$-dependencies of the scattered wave amplitude is more
than 1\%, while in the unshaded regions this error is
less than 1\%.

   If we assume that in Eqs. (14)
$\Delta x = \lambda_m/\mathrm{e}$, where
$\lambda_m = \lambda/\epsilon^{1/2}$, then the contour of
the 1\% error, obtained from Eqs. (14),
is represented by the dashed rectangle in Fig. 7.
Inside this rectangle Eqs. (14) predict
the error in the approximate scattered wave
amplitudes being less than 1\%. Outside this rectangle,
this error should be larger than 1\%. We can see
that there is a good general agreement between the
errors predicted by the approximate applicability
conditions (14) (with $\Delta x = \lambda_m/\mathrm{e}$)
and the rigorously calculated 1\% error contour
(Fig. 7).

    Very similar results can be obtained using
condition (9) that gives relative errors of the
scattered wave amplitude of $\approx 1/(2\rho_\mathrm{EAS}^2)$.
The resultant 1\% error contour for the scattered wave
amplitude is determined by the equation:
\begin{equation}
2^{1/2}\rho_\mathrm{EAS} = 10,
\end{equation}
and is presented in Fig. 7 by the thick solid curve.
Below this curve, the errors related to the
approximate theory are predicted to be less than 1\%.
This curve is again in a good agreement with the
rigorously calculated errors (Fig. 7). Note however,
that Eqs. (14) seem to provide
slightly more accurate prediction for the applicability
of the approximate theory in the case of
narrow gratings---Fig. 7. Otherwise, conditions
(9), and (10) (i.e., Eqs. (14) and (15))
are basically equivalent.

   Note again that the most interesting range of
grating amplitudes $\epsilon_1$ is where the scattered wave
amplitude is increased many times compared to
the amplitude of the incident wave. Usually, this
happens for values of $\epsilon_1$ below
$\approx 10^{-2}\epsilon$ (see Figs. 2
and 5). Fig. 7 demonstrates that in this region, the
approximate theory [29--32] is very accurate in
terms of predicting amplitudes of the scattered
wave, except for very narrow gratings with
$L \le 2.5\lambda_m$.

\section{Conclusions}

   Thus, EAS of bulk TE electromagnetic waves
in a uniform, strip like, slanted, periodic grating
with a mean dielectric permittivity that is the same
inside and outside the grating, has been rigorously
analysed in this paper. Scattering in gratings
with various grating amplitudes and grating
widths has been investigated. In particular, it has
been shown that the approximate theory, based on
the two-wave approximation and the analysis of
the diffractional divergence of the scattered wave
[29--34,37], usually gives very accurate results in
gratings with small grating amplitude, especially in
terms of predicting amplitudes of the scattered
wave.

    At the same time, it has been demonstrated that
resonantly large scattered wave amplitudes in the
geometry of EAS may result in significant effects
that cannot be explained within the approximate
theory. These are unusually strong boundary
scattering, large amplitude of the $+2$ harmonic,
and noticeable oscillations of the incident wave
amplitude in the grating. These effects become
noticeable only if the grating width is sufficiently
small (much less than the critical width [34--36]), or
if the grating amplitude exceeds $\approx 10^{-2}$ of the mean
dielectric permittivity in the structure. It has also
been demonstrated that the main source for errors
of the approximate theory of EAS in narrow
gratings with small amplitude is boundary scattering
(edge effects) at the grating interfaces.

   The applicability conditions for the approximate
theory [29--34] have been discussed, verified,
and adjusted by comparing the approximate
applicability conditions derived in paper [37] with the
results of the rigorous analysis of EAS.

\section*{Acknowledgements}

  The authors gratefully acknowledge financial
support for this research from the Queensland
University of Technology.

\section*{References}

\begin{enumerate}
\item H. Kogelnik, Bell Syst. Tech. J. 48 (1969) 2909.
\item R.S. Chu, J.A. Kong, IEEE Trans. Microwave Theory
    Tech. MTT-25 (1977) 18.
\item M.G. Moharam, T.K. Gaylord, Appl. Phys. B 28 (1982) 1.
\item T.K. Gaylord, M.G. Moharam, IEEE Proc. 73 (1985) 894.
\item E.N. Glytsis, T.K. Gaylord, J. Opt. Soc. Am. A 4 (1987)
    2061.
\item N. Chateau, J.P. Hugonin, J. Opt. Soc. Am. A 11 (1994)
    1321.
\item M.G. Moharam, E.B. Grann, D.A. Pommet, T.K.
     Gaylord, J. Opt. Soc. Am. A 12 (1995) 1068.
\item M.G. Moharam, D.A. Pommet, E.B. Grann, T.K.
     Gaylord, J. Opt. Soc. Am. A 12 (1995) 1077.
\item L. Li, J. Opt. Soc. Am. A 13 (1996) 1024.
\item J. Liu, R.T. Chen, B.M. Davies, L. Li, Appl. Opt. 38 (1999)
     6981.
\item J.M. Jarem, P.P. Banerjee, J. Opt. Soc. Am. A 16 (1999)
     1097.
\item G.I. Stegeman, D. Sarid, J.J. Burke, D.G. Hall, J. Opt.
     Soc. Am. 71 (1981) 1497.
\item E. Popov, L. Mashev, Opt. Acta 32 (1985) 265.
\item L.A. Weller-Brophy, D.G. Hall, J. Lightwave Technol. 6
     (1988) 1069.
\item D.G. Hall, Opt. Lett. 15 (1990) 619.
\item R. Petit (Ed.), Electromagnetic Theory of Gratings,
     Springer, Berlin, 1980.
\item M.C. Hutley, diffraction Gratings, Academic Press,
     London, 1982.
\item E.G. Loewen, E. Popov (Eds.), diffraction Gratings and
     Applications, Marcel Dekker, New York, 1997.
\item V.M. Agranovich, D.L. Mills (Eds.), Surface Polaritons.
     Electromagnetic Waves at Surfaces and Interfaces,
     North-Holland, Amsterdam, 1982.
\item E. Popov, L. Mashev, D. Maystre, Opt. Acta 33 (1986)
     607.
\item S.A. Akhmanov, V.I. Emel'yanov, N.I. Koroteev, V.N.
     Seminogov, Sov. Phys. Uspekhi 28 (1985) 1084.
\item V.N. Seminogov, A.I. Khudobenko, Sov. Phys. JETP 69
     (1989) 284.
\item V.I. Emel'yanov, V.I. Konov, V.N. Tokarev, V.N.
     Seminogov, J. Opt. Soc. Am. B 6 (1993) 104.
\item S. Kishino, J. Phys. Soc. Jpn. 31 (1971) 1168.
\item S. Kishino, A. Noda, K. Kohra, J. Phys. Soc. Jpn. 33
     (1972) 158.
\item T. Bedynska, Phys. Stat. Sol. (a) 19 (1973) 365.
\item T. Bedynska, Phys. Stat. Sol. (a) 25 (1974) 405.
\item A.V. Andreev, Sov. Phys. Uspekhi 28 (1985) 70 and
     references therein.
\item M.P. Bakhturin, L.A. Chernozatonskii, D.K. Gramotnev,
     Appl. Opt. 34 (1995) 2692.
\item D.K. Gramotnev, Phys. Lett. A 200 (1995) 184.
\item D.K. Gramotnev, J. Phys. D 30 (1997) 2056.
\item D.K. Gramotnev, Opt. Lett. 22 (1997) 1053.
\item D.K. Gramotnev, D.F.P. Pile, Appl. Opt. 38 (1999) 2440.
\item D.K. Gramotnev, T.A. Nieminen, J. Opt. A: Pure Appl.
     Opt. 1 (1999) 635.
\item D.K. Gramotnev, D.F.P. Pile, Phys. Lett. A 253 (1999)
     309.
\item D.K. Gramotnev, D.F.P. Pile, Opt. Quant. Electron. 32
     (2000) 1097.
\item D.K. Gramotnev, Grazing-angle scattering of
     electromagnetic waves in periodic Bragg gratings,
     to appear in Opt. Quant. Electron.
\end{enumerate}

\end{document}